\documentclass[conference]{IEEEtran}
\IEEEoverridecommandlockouts
\usepackage{cite}
\usepackage{amsmath,amssymb,amsfonts}
\usepackage{algorithmic}
\usepackage{graphicx}
\usepackage{textcomp}
\usepackage{subfigure}

\def\BibTeX{{\rm B\kern-.05em{\sc i\kern-.025em b}\kern-.08em
    T\kern-.1667em\lower.7ex\hbox{E}\kern-.125emX}}
    
\usepackage[table]{xcolor}
\usepackage{stfloats}
\usepackage{url}

\newtheorem{proposition}{Proposition}

\newcommand{\R}{\mathbb{R}}
\newcommand{\pkg}[1]{{\fontseries{b}\selectfont #1}}

\usepackage{mathrsfs}
\def\WT{\mathcal{W}}
\def\L{\mathcal L}
\def\D{D}
\def\W{W}
\def\R{\mathbb{R}}
\def\hatcov{\hat \gamma}

\definecolor{lgray}{gray}{0.75}

\definecolor{ggray}{gray}{0.85}
\def\gg{\cellcolor{ggray}}
\definecolor{agray}{gray}{0.95}

\begin{document}

\title{Large Graph Signal Denoising with\\ Application to Differential Privacy}

\author{\IEEEauthorblockN{Elie Chedemail}
\IEEEauthorblockA{
\textit{Orange Labs / CREST, ENSAI}\\
Cesson-S\'evign\'e, France \\
elie.chedemail@orange.com}
\and
\IEEEauthorblockN{Basile de Loynes}
\IEEEauthorblockA{
\textit{ENSAI}\\
Bruz, France \\
basile.deloynes@ensai.fr}
\and
\IEEEauthorblockN{Fabien Navarro}
\IEEEauthorblockA{
\textit{SAMM, Paris 1 Panth\'eon-Sorbonne}\\
Paris, France \\
fabien.navarro@univ-paris1.fr}
\and
\IEEEauthorblockN{Baptiste Olivier}
\IEEEauthorblockA{
\textit{Ericsson}\\
Stockholm, Sweden \\
baptiste.olivier@ericsson.com}
}

\maketitle

\begin{abstract}
Over the last decade, signal processing on graphs has become a very active area of research. Specifically, the number of applications, for instance in statistical or deep learning, using frames built from graphs, such as wavelets on graphs, has increased significantly. We consider in particular the case of signal denoising on graphs via a data-driven wavelet tight frame methodology. This adaptive approach is based on a threshold calibrated using Stein's unbiased risk estimate adapted to a tight-frame representation. We make it scalable to large graphs using Chebyshev-Jackson polynomial approximations, which allow fast computation of the wavelet coefficients, without the need to compute the Laplacian eigendecomposition. However, the overcomplete nature of the tight-frame, transforms a white noise into a correlated one. As a result, the covariance of the transformed noise appears in the divergence term of the SURE, thus requiring the computation and storage of the frame, which leads to an impractical calculation for large graphs. To estimate such covariance, we  develop and analyze a Monte-Carlo strategy, based on the fast transformation of zero mean and unit variance random variables. This new data-driven denoising methodology finds a natural application in differential privacy.  A comprehensive performance analysis is carried out on graphs of varying size, from real and simulated data.
\end{abstract}

\begin{IEEEkeywords}
Chebyshev polynomial approximation, Monte-Carlo methods, differential privacy, graph signal processing, Stein's unbiased risk estimate
\end{IEEEkeywords}

\section{Introduction}
Data acquired from large-scale interactive systems, such as computer, ecological, social, financial or biological networks, become increasingly widespread and accessible. In modern machine learning, the effective representation, processing or analysis of these large-scale structured data with graphs or networks are some of the key issues \cite{nickel2015review, bronstein2017geometric}. The emerging field of Graph Signal Processing (GSP) highlights connections between signal processing and spectral graph theory \cite{shuman2013emerging,ortega2018graph}, while building bridges to address these challenges. Indeed, GSP has led to numerous applications in the field of machine learning: convolutional neural networks (CNN) on graphs \cite{bruna2013spectral}, \cite{henaff2015deep,defferrard2016convolutional}, semi-supervised classification with graph CNN \cite{kipf2016semi,hamilton2017inductive} or community detection \cite{tremblay2014graph} to name just a few. We refer the reader to \cite{dong2020graph} for a recent review providing new perspectives on GSP for machine learning including, for instance, the important role it played in some of the early designs of graph neural networks (GNNs) architectures. Moreover, the recent study in \cite{fu2020understanding} shows that popular GNNs designed from a spectral perspective, such as spectral graph convolutional networks or graph attention networks, are implicitly solving graph signal denoising problems. 

In the past decades, sparse approximation with respect to a frame played a fundamental role in many areas such as signal compression and restoration, data analysis, and GSP in general. Indeed, over-complete representations like wavelet frames have several advantages and offer more flexibility over orthonormal bases. One representative family of over-complete systems derived form the orthonormal \textit{Diffusion Wavelets} of \cite{COIFMAN200653} is the so-called \textit{Spectral Graph Wavelet Transform} (SGWT) of \cite{hammond2011wavelets} constructed from a general wavelet frame. In a denoising context, SGWT has recently been adapted by \cite{gobel2018construction} to form a tight frame using the Littlewood-Paley decomposition inspired by \cite{coulhon2012heat}. Based on SGWT, \cite{de2019data} proposed an automatic calibration of the threshold parameter by adapting Stein's unbiased risk estimate (SURE) for a noisy signal defined on a graph and decomposed in a given wavelet tight frame. Even if this selection criterion produces efficient estimators of the unknown mean squared error (MSE), the main limitation is the need for a complete eigendecomposition of the Laplacian matrix, making it intractable for large-scale graphs.

We propose here to extend this methodology to large sparse graphs by avoiding this eigendecomposition, thus extending its range of application. Different strategies have been proposed in the context of GSP, one of the most popular is based on Chebyshev polynomial approximations \cite{hammond2011wavelets}. However, even if Chebyshev expansions are a good choice in many scenarios, approximations of discontinuous or non-periodic functions suffer from the \textit{Gibbs phenomenon}. A simple strategy commonly used in GSP \cite{shuman2020localized} to reduce possible spurious oscillations without additional computational cost is the introduction of Jackson's \textit{damping coefficients} \cite{jay1999, di2016efficient} which allows for higher orders of approximation.

As SURE can be evaluated in the wavelet domain, its calculation benefits directly from these efficient numerical approximations. In order to make it suitable for large sparse graphs, the only problematic step is the computation of the weights appearing in its expression. Indeed, since the SGWT is no longer orthogonal a white Gaussian noise in the graph domain is transformed into a correlated one thus involving the covariance of the transformed noise in the resulting SURE divergence term. The latter requires the explicit computation and storage of the frame in order to be calculated. Inspired by the estimation of the correlation between wavelets centered at different nodes proposed in \cite{tremblay2014graph}, our contribution is to take advantage of the interpretation of the SURE weights as the covariance between wavelet transforms of random signals in order to estimate them with Monte-Carlo approximation. We then plug this weight estimator in the SURE formula to obtain an estimator of SURE that extends well to large graph signals. In addition, we provide expressions for the variance of our proposed estimators and show that drawing Monte-Carlo samples from the centered Rademacher distribution gives a smaller variance compared to the standard Gaussian distribution. Our approach is in line with other methods \cite{ramani2008montecarlo, weller2014montecarlo} that also use Monte-Carlo strategy, but to estimate the entire divergence term involved in the calculation of SURE, in the case of uncorrelated noise.

Our proposed method can remove noise from any signal defined on a graph, this includes images \cite{shuman2013emerging} and 3D meshes \cite{onuki2016} which can have a large number of vertices. Here, we focus on an interesting application in differential privacy \cite{dwork2006} whose purpose is to protect sensitive data used by algorithms. Such privacy guarantees are usually achieved by adding white noise to the signal which inevitably reduces its statistical utility as the relevant information it contains is perturbed. This utility can be partially recovered through denoising on the condition that no information about the original signal is used. As our proposed data-driven methodology lends itself well to this usage for graph signals, we incorporate it in our numerical experiments. These give an evaluation of our Monte-Carlo estimator of SURE and its weights, along with the overall denoising methodology on both small and large graphs. In summary, the contributions of this paper are as follows:
\begin{itemize}
    \item We propose a Monte-Carlo estimation of Stein's unbiased risk estimate (SURE) that extends to signals defined on large-scale graphs. This method avoids the computationally expensive eigendecomposition of the graph Laplacian matrix required to compute weights that appear in the SURE expression.
    \item Provided expressions for the variance of our estimators show that Monte Carlo samples drawn from a Rademacher distribution is more efficient than with a Gaussian distribution. This theoretical result is illustrated through numerical experiments that compare both distributions.
    \item A performance analysis of the proposed graph signal denoising methodology shows its performance on real data protected with differential privacy and simulated large graph signals.
\end{itemize}

The paper is structured as follows. We introduce our notation of graph signals and briefly recall the SGWT definition of \cite{hammond2011wavelets}, its construction by \cite{gobel2018construction} and polynomial approximations in Section \ref{sec:denoising}. Our proposed Monte-Carlo estimators of SURE and its weights along with their respective variance are presented in Section \ref{sec:MC}. In Section \ref{sec:privacy} we present the notion of differential privacy from \cite{dwork2006} and two methods to achieve it in the context of graph signals. Finally, we numerically evaluate our estimators and compare our methodology to the DFS fused lasso introduced in \cite{padilla_2018} for small and large graphs in Section \ref{sec:simu}.

\section{Graph Signal Denoising}
\label{sec:denoising}

Consider a signal $f\in\mathbb{R}^V$ defined on an undirected weighted graph $G$, with set of vertices $V$ of cardinality $n$, and weighted adjacency matrix $\W$ with entries $(w_{ij})_{i,j\in V}$. The (unnormalized) graph Laplacian matrix $\L\in\R^{V\times V}$ associated with $G$ is the symmetric matrix defined as $\L=\D - \W$, where $\D$ is the diagonal matrix with diagonal coefficients $\D_{ii}= \sum_{j\in V} w_{ij}$. We present here our methodology with this particular Laplacian matrix but it can be easily adapted to its normalized and random walk counterparts like presented below in Section \ref{subsec:other_laplacian}.

The noise corruption model can be written as
\[
\tilde f = f + \xi,
\]
where $\xi\sim\mathcal{N}(0,\sigma^2 I_n)$. The purpose of denoising is to build an estimator $\hat f$ of $f$ that depends only on $\tilde f$.

A simple way to construct an effective non-linear estimator is obtained by thresholding the SGWT coefficients of $f$ on a frame (see \cite{hammond2011wavelets} for details about the SGWT). Given the Laplacian and a given frame, denoising in this framework can be summarized as follows:
\begin{itemize}
\item Analysis: compute the SGWT transform $\WT \tilde f$;
\item Thresholding: apply a given thresholding operator (\emph{e.g.,} \textit{soft} or \textit{hard}) to the coefficients $\WT  \tilde f$; 
\item Synthesis: apply the inverse SGWT transform to obtain an estimate $\hat f$ of the original signal.
\end{itemize}

This procedure can be viewed as an extension of the wavelet denoising methodology from Donoho and Johnstone \cite{donoho1995} to the SGWT.

\subsection{Spectral Graph Wavelet Transform}
\label{subsec:sgwt}

The SGWT decomposes a signal into a frame $\mathfrak F=\{ r_i \}_{i \in I}$ of vectors of $\mathbb R^V$ with frame bounds $A$, $B > 0$ satisfying for all $f \in \mathbb R^V$
\begin{equation*}
A\|f \|^2_2 \leq \sum_{i \in I} |\langle f,r_i \rangle|^2 \leq B \|f\|^2_2.  
\end{equation*}
When $A = B = 1$, the above inequality becomes Parseval's identity and such a frame is said to be \textit{tight}.

As $\L$ is a symmetric matrix, its spectral decomposition is given by $\L=\sum_{\ell} \lambda_\ell \langle \chi_\ell,\cdot \rangle \chi_\ell$, where $\lambda_{1} \geq \lambda_{2} \geq \cdots \geq \lambda_{n} = 0$ are the (ordered) eigenvalues of $\L$ and $(\chi_{\ell})_{1\leq \ell\leq n}$ are the associated eigenvectors. Then for any function $\rho:\mathrm{sp}(\L)\rightarrow \R$ defined on the spectrum of $\L$, we have the functional calculus formula $\rho(\L) = \sum_{\ell} \rho(\lambda_{\ell}) \langle \chi_{\ell}, \cdot \rangle \chi_{\ell}$.

\begin{figure}[h]
    \centering
    \includegraphics[width=0.48\textwidth]{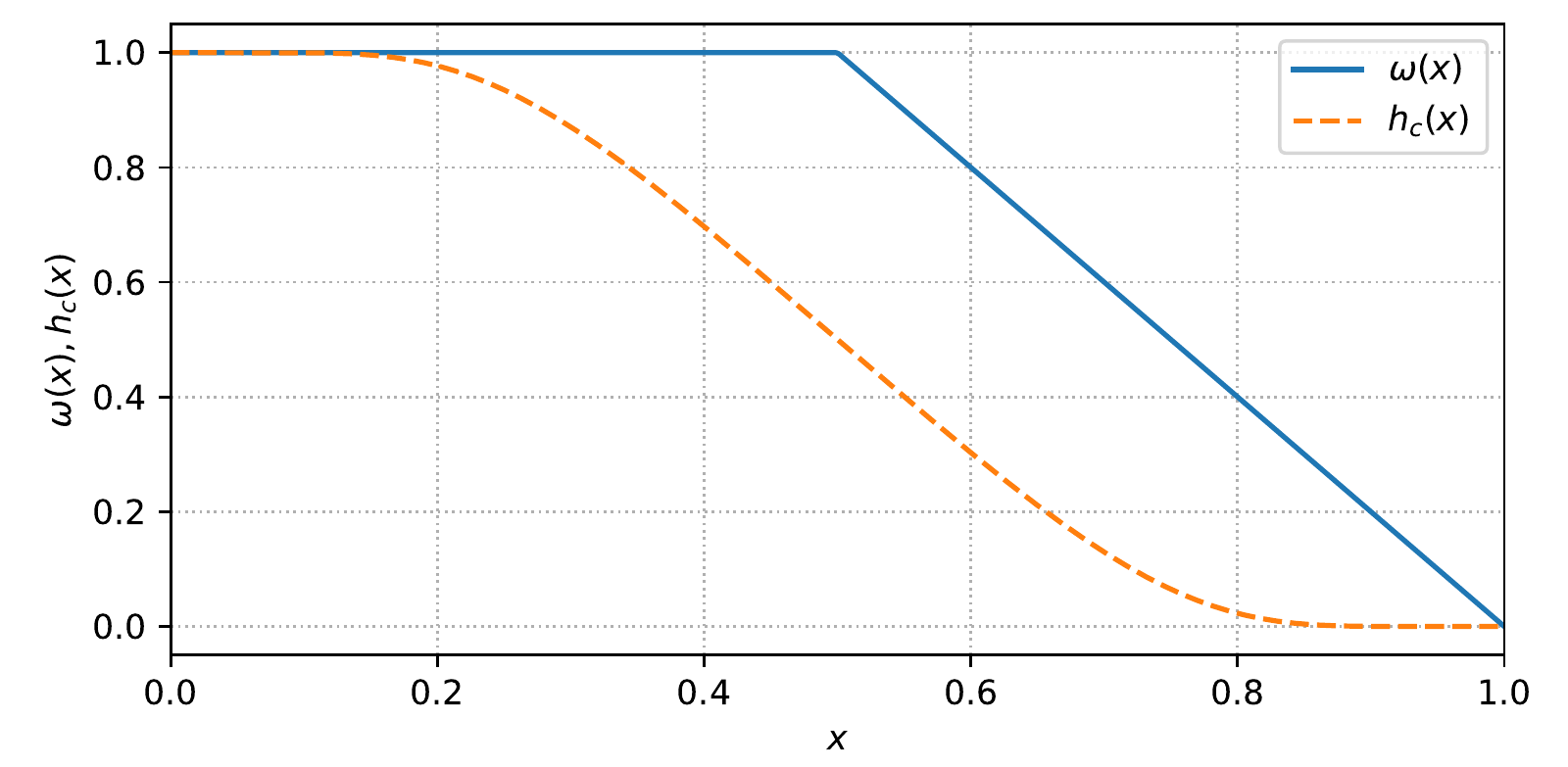}
    \caption{$\omega$ and $h_c$ functions on $[0, 1]$}
    \label{fig:omega_gc}
\end{figure}

We build a tight frame following \cite{Leonardi2013, gobel2018construction} with a finite partition of unity $(\psi_j)_{j=0, \ldots,J}$ on the compact $[0,\lambda_1]$ defined as follows: let $\omega : \mathbb R^+ \rightarrow [0,1]$ be some continuous function with support in~$[0,1]$, satisfying $\omega \equiv 1$ on $[0,b^{-1}]$, for some $b>1$, and set
\begin{equation*}
\psi_0(x)=\omega(x),~~\psi_j(x)=\omega(b^{-j}x)-\omega(b^{-j+1}x),
\end{equation*}
for $j=1, \ldots, J$, where $J= \lfloor \log \lambda_1 / \log b \rfloor + 2$. In our numerical experiments, we use the following piecewise linear function $\omega$:
\begin{equation*}
  \omega(x) =
    \begin{cases}
      1 & \text{if } 0 \leq x \leq b^{-1}\\
      \frac{b}{1-b} x + \frac{b}{b-1} & \text{if } b^{-1} < x \leq 1\\
      0 & \text{if } x > 1
    \end{cases},
\end{equation*}
with parameter $b=2$. An alternative $\omega$ function is the $\mathcal C^\infty$ function $h_c(x) = g_c(x+1) g_c(1-x)$ where $g_c(x) = f(x)/(f(x) + f(c-x))$ and $f(x) = e^{-1/x} \mathbf 1_{\{x > 0\}}$. Another choice is to take a $\mathcal C^3$ piecewise polynomial plateau function like the authors of \cite{gobel2018construction}. Figure \ref{fig:omega_gc} illustrates the $\omega$ and $h_c$ functions with parameters $b=2$ and $c=1$, respectively. 

\begin{figure}[h]
    \centering
    \includegraphics[width=0.48\textwidth]{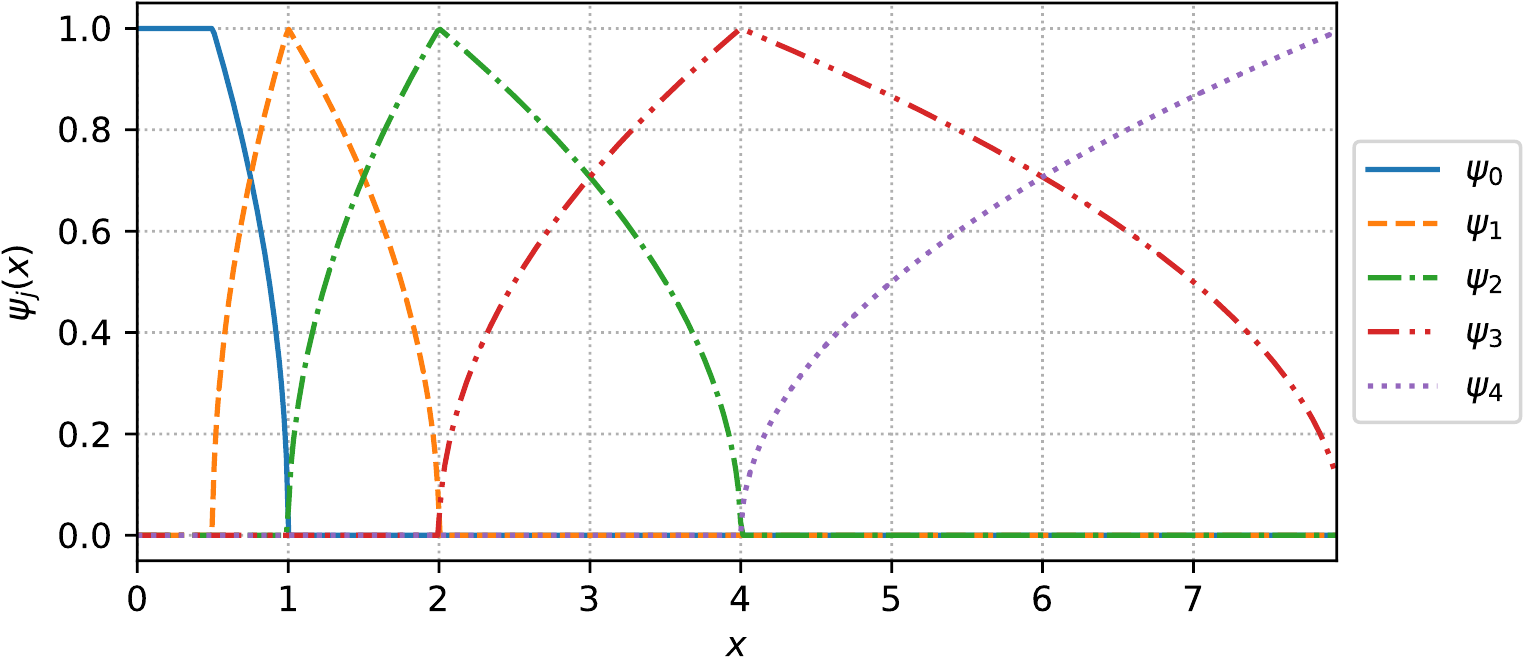}
    \caption{Finite partition of unity on $[0,\lambda_1]$}
    \label{fig:psi_j}
\end{figure}

The partition of unity $(\psi_j)_{j=0, \ldots,4}$ on $[0,\lambda_1]$ obtained with the graph presented in the first experiment from Section \ref{sec:simu} is shown in Figure \ref{fig:psi_j}.

Using Parseval's identity, we can show that the following set is a tight frame:
\begin{equation*}
    \mathfrak F = \left \{ \sqrt{\psi_j}(\L)\delta_i, j=0, \ldots, J, i \in V \right \}.
\end{equation*}
Decomposing a signal $f \in \mathbb R^V$ into this frame results in its SGWT along the $(J+1)$ scales:
\begin{equation*}
    \WT f = \left ( \sqrt{\psi_0}(\L)f^{T},\ldots,\sqrt{\psi_J}(\L)f^{T} \right )^{T} \in \mathbb R^{n(J+1)}.
\end{equation*}
With the tightness property of the frame, the inverse transform is directly given by the application of the adjoint matrix to the wavelet coefficients:
\begin{equation*}
    \WT^\ast \left (\eta_{0}^{T}, \eta_{1}^{T}, \ldots, \eta_{J}^T \right )^{T} = \sum_{j\geq 0} \sqrt{\psi_j}(\L)\eta_{j}.
\end{equation*}

\subsection{SGWT Polynomial Approximation}

Direct computation of the SGWT entails functional calculus on the graph Laplacian matrix $\L$ and thus the computation of its eigenvectors and eigenvalues. This limits applications to reasonably sized graphs that have less than a few thousand nodes. For larger ones, the computationally expensive eigendecomposition can be avoided through a fast transform based on Chebyshev polynomial approximation \cite{hammond2011wavelets}.

The Chebyshev polynomials of the first kind $T_k(x)$ are obtained from the recurrence relation $T_k(x)=2xT_{k-1}(x)-T_{k-2}(x)$, with $T_0(x)=1$ and $T_1(x)=x$. They form an orthogonal basis of the Hilbert space $\mathbb{L}^2([-1, 1], dy/\sqrt{1-y^2})$. Any filter $\rho$ can be approximated with the truncated Chebyshev expansion of degree $K$
\begin{equation*}
    \rho_K(\L) =\sum_{i=0}^{K}\theta_i(\tilde \rho)T_i(\widetilde \L),
\end{equation*}
where $\theta_i(\tilde \rho)$ is the $i$-th coefficient of the Chebyshev expansion of function $\tilde \rho (x) = \rho(\frac{\lambda_1}{2}(x+1))$ and $T_i(\widetilde \L)$ is the $i$-degree Chebyshev polynomial computed for $\widetilde \L = \frac{2}{\lambda_1} \L - I_n$. This transformation of $\L$ extends the expansion to any Laplacian matrix by mapping $[0, \lambda_1]$ into $[-1, 1]$. According to \cite{hammond2011wavelets}, for all filter $\rho$ defined on $\mathrm{sp}(\L)$ and all signal $f$, the approximation $\rho_K(\L) f$ is close to $\rho(\L)f$. 

While this first approximation is more practical than the complete SGWT, it is subjected to the \textit{Gibbs phenomenon}. A solution is to include Jackson coefficients $g_i^K$ as damping multipliers in the Chebyshev expansion:
\begin{equation*}
    \rho_K(\L) =\sum_{i=0}^{K} g_i^K \theta_i(\tilde \rho)T_i(\widetilde \L).
\end{equation*}
An expression of these damping factors can be found in \cite{jay1999}, a shorter form proposed in \cite{di2016efficient} is given by
\begin{equation*}
    g_i^K = \frac{\sin(i+1)\alpha_K}{(K+2)\sin(\alpha_K)} + \left( 1 - \frac{i+1}{K+2} \right) \cos(i\alpha_K),
\end{equation*}
where $\alpha_K = \pi / (K+2)$. This Chebyshev-Jackson polynomial approximation reduces Gibbs oscillations resulting in a better convergence as the degree $K$ increases.

\subsection{Extension to Other Laplacian Matrices}
\label{subsec:other_laplacian}

As previously mentioned, this methodology also adapts well to the normalized and random walk (or asymmetric) Laplacian matrices, respectively defined as $\L_{\rm norm} = D^{-\frac{1}{2}} \L D^{-\frac{1}{2}}$ and $\L_{\rm rw} = D^{-1} \L$. These have been used as an alternative to the unnormalized graph Laplacian $\L$ in other related methods such as the graph Fourier transform \cite{girault2018}.

The normalized Laplacian matrix is real symmetric like $\L$ which means it is diagonalizable and therefore suited for our approach. As its spectrum $\mathrm{sp}(\L_{\rm norm}) = \{\mu_1, \dots, \mu_n\}$ is always contained in the interval $[0, 2]$, its maximum eigenvalue $\mu_1$ is bounded by 2. This represents a special case of the construction described above and requires a few modifications. First, the formula $J = \lfloor \log \mu_1 / \log b \rfloor + 2$ that determines the number of scales in the wavelet decomposition restricts the choice of parameter $b$ to the interval $(1, 2]$ in order to get more than $J + 1 = 3$ scales. Then, the polynomial approximation consists of the truncated Chebyshev expansion of function $\tilde \rho (x) = \rho (x+1)$ with the appropriate transformation $\widetilde \L_{\rm norm} = \L_{\rm norm} - I_n$.

On the other hand, the random walk Laplacian matrix is not symmetric but is diagonalizable nonetheless as it is similar to the normalized Laplacian: $\L_{\rm norm} = D^{\frac{1}{2}} \L_{\rm rw} D^{-\frac{1}{2}}$. Its eigenvalues and eigenvectors are easily obtained from the eigendecomposition of $\L_{\rm norm}$:
\begin{align*}
    \L_{\rm norm} u_\ell &= \mu_\ell u_\ell \\
    D^{\frac{1}{2}} \L_{\rm rw} D^{-\frac{1}{2}} u_\ell &= \mu_\ell u_\ell \\
    \L_{\rm rw} (D^{-\frac{1}{2}} u_\ell) &= \mu_\ell (D^{-\frac{1}{2}} u_\ell),
\end{align*}
where $u_\ell$ is the eigenvector of $\L_{\rm norm}$ associated with $\mu_\ell$. We see that $\L_{\rm rw}$ has exactly the same spectrum as $\L_{\rm norm}$ and its set of eigenvectors is given by $\{D^{-\frac{1}{2}} u_\ell\}_{\ell=1, \dots, n}$. Since these form an orthonormal basis for $\mathbb R^n$ with the inner product $\langle x, y \rangle_D = x^\top D y$, we have the spectral decomposition $\L_{\rm rw}=\sum_{\ell} \mu_\ell \langle D^{-\frac{1}{2}} u_\ell, \cdot \rangle_D D^{-\frac{1}{2}} u_\ell$. The functional calculus formula for any function $\rho$ defined on the spectrum of $\L_{\rm rw}$ is thus $\rho(\L_{\rm rw})=\sum_{\ell} \rho(\mu_\ell) \langle D^{-\frac{1}{2}} u_\ell, \cdot \rangle_D D^{-\frac{1}{2}} u_\ell$. Chebyshev polynomial approximation can then be applied in the same way as for the normalized Laplacian matrix.

\section{Monte-Carlo Estimation of Weights}
\label{sec:MC}

From \cite{de2019data}, SURE for a general thresholding process $h:\mathbb{R}^{n(J+1)} \rightarrow \mathbb{R}^{n(J+1)}$ is given by the following identity  
\begin{equation} \label{eq:sureh}
\mathbf{SURE}(h)=-n \sigma^2 + \|h(\widetilde F)-\widetilde F\|^2 + 2 \sigma^2 \sum_{i,j=1}^{n(J+1)} \gamma_{ij}^2 \partial_j h_i(\widetilde F),
\end{equation}
where $\widetilde F = \WT \tilde f$ is the wavelet transform of the noisy signal $\tilde f$. In \cite{de2019data}, the weights $\gamma_{ij}^2 = (\WT \WT^*)_{ij}$, $i,j=1, \ldots, n(J+1)$, are computed from the full reduction of the Laplacian matrix which is no longer tractable for large graphs. However, as shown in \cite{hammond2011wavelets}, the SGWT can be efficiently approximated by using Chebyshev polynomials. Besides, it is clear from the probabilistic interpretation given in \cite[Th. 1]{de2019data} that
\begin{equation*}
  \forall i,j=1, \ldots, n(J+1), \quad \gamma_{ij}^2=\mathbf E[(\WT \varepsilon)_i (\WT \varepsilon)_j ].
\end{equation*}
where $\varepsilon=(\varepsilon_1, \ldots, \varepsilon_n)$ are \emph{i.i.d.} random variables with zero mean and variance one. Thus, taking advantage of this identity, we propose to estimate the weights with Monte-Carlo approximation as follows:
\begin{itemize}
\item generate $(\varepsilon_{ik})_{i=1, \ldots, n, k=1, \ldots, N}$ of \emph{i.i.d.} random variables such that $\mathbf E[\varepsilon_{ik}]=0$ and $\mathbf V(\varepsilon_{ik})=1$;
\item compute
\begin{equation*}
    \hatcov^2_{ij} = \frac{1}{N} \sum_{k=1}^N (\WT \varepsilon_k)_i (\WT \varepsilon_k)_j,
\end{equation*}
where $\varepsilon_k = (\varepsilon_{ik})_{i=1, \ldots, n}$ are random signals.
\end{itemize}

Generally speaking, whereas Monte-Carlo are simple methods to implement and can be easily parallelized, they suffer from their slow rate of convergence. In practice, a well-chosen distribution for the random variables $\varepsilon_{ik}$ can result in a lower variance of the estimator $\mathbf V[\hatcov_{ij}^2]$ whose expression is given below. In fact, it is even more interesting to compute the variance of SURE when the estimator $\hatcov^2_{ij}$ is plugged in place of the weights $\gamma_{ij}^2$ in~\eqref{eq:sureh}.

\subsection{Variance of $\hatcov^2_{ij}$.}
\label{subsec:gamma_variance}

A straightforward computation gives the expectation of $\hatcov^2_{ij}$
\begin{align*}
  \mathbf E[\hatcov^2_{ij}] & = \mathbf E \left [ \left ( \sum_{p=1}^n \WT_{ip} \varepsilon_{p1} \right ) \left ( \sum_{p=1}^n \WT_{jp} \varepsilon_{p1} \right ) \right ] \\
  & = \sum_{p,q=1}^n \WT_{ip} \WT_{jq} \mathbf E[\varepsilon_{p1} \varepsilon_{q1}] = \sum_{p=1}^n \WT_{ip} \WT_{jp} = \gamma_{ij}^2.
\end{align*}
The variance of $\hatcov^2_{ij}$ is given by the following result and its computation is derived in Appendix \ref{appendix:gamma_variance}.
\begin{proposition}
\begin{align*}
    \mathbf V [\hatcov^2_{ij}] & = \frac{1}{N} \Bigg\{ \mathbf V[\varepsilon_{11}^2] \sum_{p=1}^n \WT_{ip}^2 \WT_{jp}^2 \\
    & + 2\mathbf E[\varepsilon_{11}^2]^2 \sum_{\substack{p,q=1,\\ p \neq q}}^n \WT_{ip} \WT_{iq} \WT_{jp} \WT_{jq} \Bigg\}.
\end{align*}
\end{proposition}
Note the usual rate of convergence $\sqrt{N}$ from Monte-Carlo estimation. In many papers in the literature $\varepsilon_{11}$ is chosen to be distributed as a standard Gaussian random variable so that $\mathbf V[\varepsilon_{11}^2]=2$. However, if $\varepsilon_{11}$ has a centered Rademacher distribution with probability mass function $\frac{1}{2} \delta_{-1} + \frac{1}{2} \delta_1$, then $\varepsilon_{11}^2$ is deterministic and $\mathbf V[\varepsilon_{11}^2]=0$. With such a choice, the variance of $\hatcov_{ij}^2$ is then reduced to
\begin{equation*}
\mathbf V[\hatcov_{ij}^2]= \frac{2}{N} \sum_{\substack{p,q=1, \\ p \neq q}}^n \WT_{ip} \WT_{iq} \WT_{jp} \WT_{jq}.
\end{equation*}
This trick is actually well known in the literature \cite{Hutchinson1990}. This computation somehow provides arguments in favor of the Rademacher distribution.

Another way to further reduce the variance of $\hatcov_{ij}^2$ is to take advantage of the SGWT localization property. Let us denote by $\lfloor x \rfloor$ the integer part of a real $x \in \mathbb R$. Then, for any $i \in \{ 1, \ldots, n(J+1) \}$ and any $p \in \{1, \ldots, n \}$
\begin{equation*}
  |\WT_{ip}| = \left | \left \langle \sqrt{\psi_{\lfloor i/n \rfloor}(\L)} \delta_{i-\lfloor i/n \rfloor},\delta_p \right \rangle \right | \leq \| \psi_{\lfloor i/n \rfloor}(\L) \|_2^2 \leq 1.
\end{equation*}
Since the SGWT is localized both in the space and the frequency domain, $\WT_{ip}$ vanishes as the geodesic distance between $i-\lfloor i/n \rfloor$ and $p$ grows. Thus, the performance of the Monte-Carlo estimation could be improved by a suitable calibration of the partition of unity. As a consequence, most terms in the expression of $\mathbf V[\hatcov_{ij}^2]$ are small thanks to the localization properties of SGWT.

\subsection{Variance of $\mathbf{\widehat{SURE}}$}
\label{subsec:sure_variance}

The SURE plug-in estimator is obtained by replacing the weights $\gamma_{ij}^2$ with their Monte-Carlo estimators $\hatcov^2_{ij}$:
\begin{equation*}
    \mathbf{\widehat{SURE}}(h)=-n \sigma^2 + \|h(\widetilde F)-\widetilde F\|^2 + 2 \sigma^2 \sum_{i,j=1}^{n(J+1)} \hatcov_{ij}^2 \partial_j h_i(\widetilde F).
\end{equation*}
Given the observed wavelet coefficients $\widetilde F$, this estimator of SURE has no bias as it is a linear function of the unbiased weight estimators $\hatcov^2_{ij}$. The following proposition presents its conditional variance whose computation is detailed in Appendix \ref{appendix:sure_variance}.
\begin{proposition}
\begin{align*}
  \mathbf V[\mathbf{\widehat{SURE}}(h) | \widetilde F] & = \frac{4 \sigma^4}{N} \sum_{i,j,k,\ell=1}^{n(J+1)} \partial_j h_i(\widetilde F) \partial_k h_\ell(\widetilde F) \Bigg \{ \\
  & \phantom{+~} \mathbf V[\varepsilon_{11}^2] \sum_{p=1}^n \WT_{ip} \WT_{jp} \WT_{kp} \WT_{\ell p} \\
  & + \mathbf E[\varepsilon_{11}^2]^2 \sum_{\substack{p,q=1, \\ p \neq q}}^{n} \WT_{ip} \WT_{jq} \WT_{kp}\WT_{\ell q} \\
  & + \mathbf E[\varepsilon_{11}^2]^2 \sum_{\substack{p,q=1, \\ p \neq q}}^{n} \WT_{ip} \WT_{jq} \WT_{kq}\WT_{\ell p} \Bigg \}.
\end{align*}
\end{proposition}
Here again, the Rademacher distribution reduces the variance compared to the Gaussian distribution. 

\subsection{Computational Complexity}

The polynomial approximation of all random signal wavelet transforms $\mathcal W \varepsilon_k$ is of order $\mathcal O (N(m K + n(J+1)K))$, where $m$ is the number of edges in the graph \cite{hammond2011wavelets}. Then, computing every $(\hatcov^2_{ij})_{i,j=1,\dots,n(J+1)}$ term requires $\mathcal O(n^2 (J+1)^2 (2N-1))$ operations. The computation of all the weights is useful when performing block thresholding on the wavelet coefficients $\mathcal W \tilde f$ \cite{de2019data} which shows good denoising performance but is relatively computationally expensive. Alternatively, a coordinate-wise thresholding process only needs the diagonal weights $(\hatcov^2_{ii})_{i=1,\dots,n(J+1)}$ whose computation is reduced to $\mathcal O(n(J+1)N)$ operations.

After this initial weight estimation, the computational complexity for the approximated wavelet transform of the noisy signal $\tilde f$ is $\mathcal O (m K + n(J+1)K)$. The coordinate-wise thresholding step has an average cost of $\mathcal O (n(J+1) \log(n(J+1)))$ according to \cite{donoho1995}. Finally, the approximated inverse transform has the same complexity as the forward transform.

\section{Differential Privacy and Gaussian Mechanism}
\label{sec:privacy}

We now give a definition of differential privacy \cite{dwork2006} which constitutes a strong standard for privacy guarantees about algorithms that use sensitive data. Let $X_1, \dots, X_m$ be a random vector containing the private data of $m$ individuals we wish to protect with a \textit{privacy mechanism}. This information is collected in a dataset $X = (X_1, \dots, X_m)$ that serves as an input to the mechanism which returns a \textit{sanitized} output $Z = (Z_1, \dots, Z_k)$ that preserves the privacy of each individual. Let $(\mathcal X^m, \mathcal A^m)$ and $(\mathcal Z, \mathcal B)$ be the measurable spaces where $X$ and $Z$ respectively take values. A privacy mechanism $Q(\cdot | X)$ corresponds to the conditional distribution of $Z$ given $X$, that is $Q(A | x) = \mathbb{P}(Z \in A | X = x)$, where $Q(\cdot | \cdot) : \mathcal B \times \mathcal X^m \rightarrow [0, 1]$ is a Markov kernel.

Let $\varepsilon \geq 0$ be the privacy budget and $\delta \geq 0$ another privacy parameter. The privacy mechanism $Q$ is said to satisfy $(\varepsilon, \delta)$-differential privacy if for any two datasets $x, x^\prime \in \mathcal X^m$ that differ on a single entry and for any subset of outputs $A \in \mathcal B$, we have
\begin{equation*}
    Q(A | x) \leq e^\varepsilon Q(A | x^\prime) + \delta.
\end{equation*}
As it appears from this definition, smaller privacy parameters lead to closer output distributions and hence a better privacy preserving mechanism. Intuitively, differential privacy protects individuals by ensuring the inclusion or removal of their information from the input dataset does not affect much the output distribution.

In this paper, we are interested in functions $f : \mathcal X^m \rightarrow \mathbb R^n$ that map a dataset $X$ to a graph signal $f \in \mathbb R^n$. In order to achieve differential privacy, a common method is to introduce just enough uncertainty in the function response to hide the participation of any single individual. The Gaussian mechanism does so by adding white Gaussian noise $\xi \sim \mathcal{N} (0, \sigma^2 I_n)$ to the response where the standard deviation $\sigma$ is calibrated to the privacy parameters and a third quantity $\Delta$ called the $l_2$-sensitivity. It is defined by $\Delta = \max \| f(x) - f(x^\prime) \|_2$ which corresponds to the maximum impact a single individual's information can have on the signal $f$. This method yields a sanitized output $Z = f(X) + \xi$ that can be interpreted as a noisy signal $\tilde f = f + \xi$ with our noise corruption model given in Section \ref{sec:denoising}.

We present two Gaussian mechanisms that use different variance formulas to sanitize a function $f : \mathcal X^m \rightarrow \mathbb R^n$ with $l_2$-sensitivity $\Delta$. First, the classical Gaussian mechanism proposed by \cite{dwork2006} preserves ($\varepsilon$, $\delta$)-differential privacy for any $\varepsilon$, $\delta \in (0, 1)$ if $\sigma \geq \Delta \sqrt{2 \log(1.25 / \delta)} / \varepsilon$. The authors of \cite{balle2018} have shown that this formula is not optimal and can be further improved to reduce the amount of noise needed to achieve the same degree of privacy. Whereas the classical Gaussian mechanism uses a Gaussian tail approximation to obtain a bound for the standard deviation, their proposed approach uses numerical evaluations of the cumulative Gaussian distribution function $\Phi (x) = \mathbb{P}(\mathcal{N}(0, 1) \leq x)$ to determine an optimal variance. Their analytic Gaussian mechanism preserves ($\varepsilon$, $\delta$)-differential privacy for any $\varepsilon \geq 0$ and $\delta \in [0, 1]$ if and only if
\begin{equation*}
    \Phi \left( \frac{\Delta}{2 \sigma} - \frac{\varepsilon \sigma}{\Delta} \right) - e^\varepsilon \Phi \left( - \frac{\Delta}{2 \sigma} - \frac{\varepsilon \sigma}{\Delta} \right) \leq \delta.
\end{equation*}
The Gaussian mechanism offers a solution to sanitize a signal at the expense of its utility as it is perturbed by the introduced noise. Indeed, there is a trade-off between privacy and utility: very small values of $\varepsilon$ and $\delta$ ensure a strong degree of privacy but can be detrimental to the utility of the sanitized data, and vice versa. A valuable aspect of differential privacy is its immunity to post-processing \cite[Prop. 2.1]{dwork2014} as long as no knowledge about the original signal is used. Formally, the composition of any data-independent function with an $(\varepsilon, \delta)$-differentially private mechanism is also $(\varepsilon, \delta)$-differentially private. Therefore, a data-driven denoising method such as ours can improve the utility of a sanitized signal with no loss of privacy. Indeed, the SGWT, SURE and their respective approximations only require information contained in the known Laplacian matrix $\L$ and observed noisy signal $\tilde f$.

\section{Numerical Experiments}
\label{sec:simu}

In this section, we present an experimental evaluation of our Monte-Carlo estimator of the SURE weights described in Section \ref{sec:MC} and graph signal denoising methodology. Specifically, we are interested in the sanitization and denoising of density maps of located events over a given period. First, we consider signals built from real datasets gathering positions of taxis in the cities of New York and San Francisco. The relatively small size of their corresponding graphs enables us to diagonalize their respective Laplacian matrices and directly compute the SGWT and SURE in a reasonable amount of time, thus allowing the comparison with our approximation method. Then, we generate signals on a large graph for which the eigendecomposition of the Laplacian matrix is not tractable. Hereafter, wavelet transforms are performed with the piecewise linear $\omega$ function presented in Section \ref{subsec:sgwt} and each polynomial approximation is of degree $K = 100$.

\subsection{Monte-Carlo Estimator of the Weights $\gamma_{ii}^2$}

This experiment makes use of the New York City (NYC) yellow taxi trip records publicly released each year by the Taxi and Limousine Commission (TLC). These datasets contain in particular the pickup and drop-off locations and times of each trip whose distribution has been the subject of different GSP applications \cite{ortega2018graph, balle2018, chen2018multiresolution}. In the past, a bad pseudonymization of the taxi ID led to a privacy breach \cite{douriez2016} for the drivers and their passengers about where they might reside and the places they frequent. 

Since the yellow taxis mostly operate in the borough of Manhattan, we focus our experiment on Manhattan Island and build an associated graph with \pkg{OSMnx} \cite{boeing2017}. This Python package automatically downloads urban networks from the OpenStreetMap database, converts them to graph objects of the \pkg{NetworkX} package and offers a variety of analysis tools. The resulting graph consists of 4513 nodes and 9743 edges representing street intersections and segments, respectively.

With this first graph, we evaluate the SURE weights Monte-Carlo estimators when their samples are either drawn from a centered Rademacher or standard Gaussian distribution. Their SGWT are computed with the Chebyshev polynomial approximation and we estimate the weights for different Monte-Carlo sample sizes $N$. We focus on the diagonal weights $\hat{\gamma}_{ii}^2$, $i=1, \ldots, n(J+1)$ as only these are needed in the coordinate-wise thresholding process and compare them to the weights obtained from the complete transform by averaging the MSE along the $n$ nodes and $(J+1)$ scales over 50 repetitions: $\textrm{MSE}((\gamma_{ii}^2)_i, (\hat{\gamma}_{ii}^2)_i) = \frac{1}{n(J+1)} \sum_{i=1}^{n(J+1)} (\gamma_{ii}^2 - \hat{\gamma}_{ii}^2)^2$.

\begin{figure}[ht]
    \centering
    \includegraphics[width=0.48\textwidth]{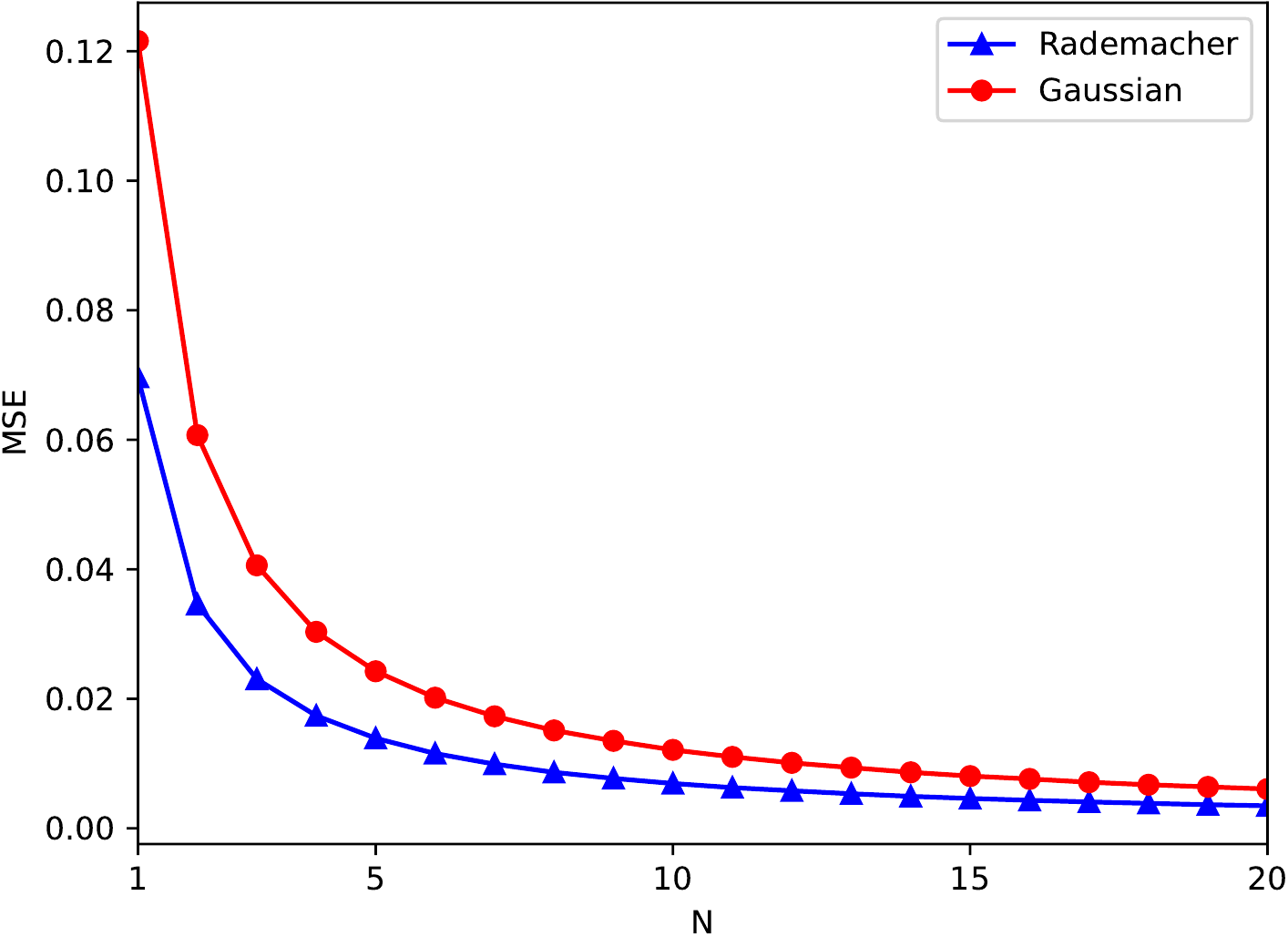}
    \caption{Average MSE between the SURE weights $\gamma_{ii}^2$ and their Monte-Carlo estimators $\hat{\gamma}_{ii}^2$ over 50 repetitions.}
    \label{fig:gamma_ii}
\end{figure}

Results are presented in Figure \ref{fig:gamma_ii} where we see that drawing samples from the centered Rademacher distribution gives estimates closer to the real weights in terms of MSE for any sample size compared to the standard Gaussian distribution. This illustrates the gain in variance achieved with the former distribution as mentioned in Section \ref{subsec:gamma_variance}.

\subsection{SURE Monte-Carlo Estimator}

We now present how SURE behaves when the estimated weights $\hat{\gamma}_{ii}^2$ are plugged in. On the same graph, we build a signal $f$ by counting the number of taxi pickups projected to the nearest intersection over a period of one hour. Here, we consider the time interval between 00:00 and 01:00 on Sept 24, 2014, as chosen by \cite{balle2018} to compare our results in a similar configuration. We add some white Gaussian noise with standard deviation $\sigma=1$ to obtain a noisy signal and compute its SGWT coefficients with the Chebyshev-Jackson approximation. The denoising is done by applying the James-Stein thresholding function $\tau(x, t) = x \max \{ 1 - t^2 |x|^{-2}, 0 \}$ to the coefficients with the threshold that minimizes $\textrm{MSE}(f, \hat{f})$. Finally, for this optimal threshold we estimate SURE between the estimate $\hat{f}$ and the original signal with known $\sigma$ for different Monte-Carlo sample sizes and both the centered Rademacher and standard Gaussian distributions.

\begin{figure}[ht]
    \centering
    \includegraphics[width=0.48\textwidth]{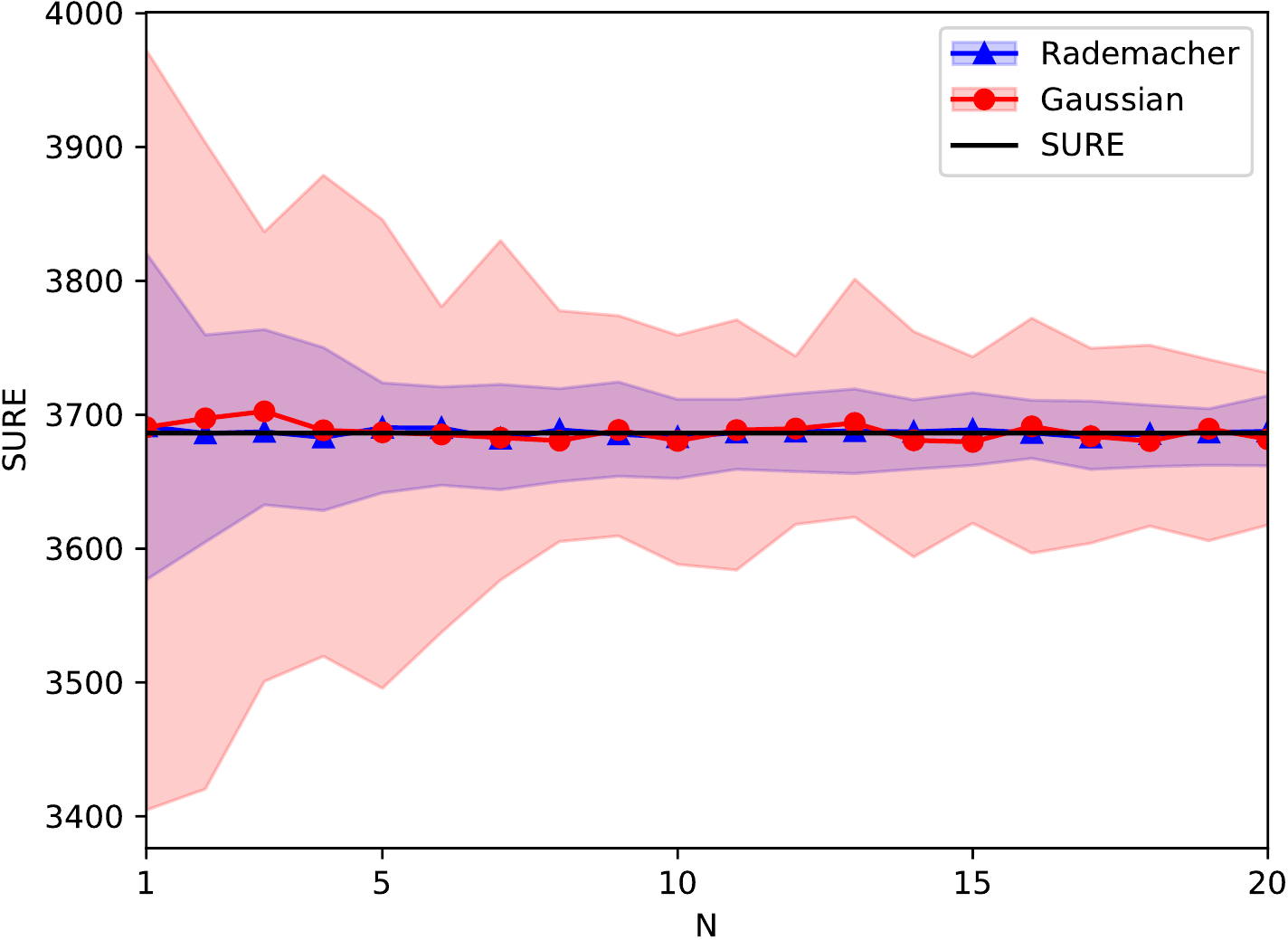}
    \caption{Average SURE Monte-Carlo estimate and 95\% CI over 50 repetitions.}
    \label{fig:sure_mc}
\end{figure}

Figure \ref{fig:sure_mc} shows the average SURE estimate over 50 repetitions along with a 95 \% empirical confidence interval. We visualize the SURE plug-in estimator unbiasedness as it is centered on the real SURE value. Additionally, we observe the smaller variance of the Monte-Carlo estimator when samples are drawn from the centered Rademacher distribution compared to the Gaussian distribution, as previously mentioned in Section \ref{subsec:sure_variance}. As a result, we estimate SURE in the following experiments with 10 samples from this distribution.

\subsection{Denoising of Differentially Private Graph Signals}

\begin{table*}[ht]
\centering
\caption{Average SNR performance over 10 realizations of high to low privacy budget sanitization on the NYC graph.}
\label{table:ny}
\begin{tabular}{lcccccccc}
\hline
                  & \multicolumn{4}{c}{Classical Gaussian mechanism}                                                                                                                     & \multicolumn{4}{c}{Analytic Gaussian mechanism}                                                                                                                       \\ \hline
$\varepsilon$           & 0.2                                    & 0.3                                    & 0.5                                    & 1                                      & 0.2                                    & 0.3                                    & 0.5                                    & 1                                       \\
$\sigma$             & 26.49                                   & 17.66                                   & 10.60                                   & 5.30                                   & 18.99                                   & 12.99                                   & 8.06                                    & 4.22                                    \\
$\textrm{SNR}_{\textrm{in}}$           & -12.48 $\pm$ 0.11                       & -8.96 $\pm$ 0.11                        & -4.52 $\pm$ 0.11                        & 1.5 $\pm$ 0.11                         & -9.59 $\pm$ 0.11                        & -6.29 $\pm$ 0.11                        & -2.14 $\pm$ 0.11                        & 3.47 $\pm$ 0.11                         \\ \hline
$\textrm{SGWT}_{\textrm{MSE}}$        & 0.23 $\pm$ 0.32                         &\gg 1.61 $\pm$ 0.26 & \gg 3.55 $\pm$ 0.23 & \gg 7.0 $\pm$ 0.16 & \gg 1.36 $\pm$ 0.27 & \gg 2.73 $\pm$ 0.23 &\gg 4.79 $\pm$ 0.18 & \gg 8.29 $\pm$ 0.15 \\
$\textrm{SGWT}_{\textrm{SURE}}$       & 0.1 $\pm$ 0.24                          & 1.5 $\pm$ 0.31                          & 3.52 $\pm$ 0.2                          & 6.94 $\pm$ 0.17                        & 1.25 $\pm$ 0.38                         & 2.71 $\pm$ 0.25                         & 4.75 $\pm$ 0.2                          & 8.24 $\pm$ 0.18                         \\
$\textrm{SGWT}_{\textrm{SURE, MC}}^{\textrm{CJ}}$ & 0.1 $\pm$ 0.28                          & 1.51 $\pm$ 0.34                         & 3.52 $\pm$ 0.24                         & 6.92 $\pm$ 0.15                        & 1.3 $\pm$ 0.34                          & 2.71 $\pm$ 0.25                         & 4.74 $\pm$ 0.22                         & 8.24 $\pm$ 0.17                         \\
$\textrm{DFS}_{\textrm{MSE}}$         & \gg 0.88 $\pm$ 0.02 & 1.18 $\pm$ 0.09                         & 2.33 $\pm$ 0.18                         & 5.58 $\pm$ 0.15                        & 1.09 $\pm$ 0.09                         & 1.72 $\pm$ 0.12                         & 3.43 $\pm$ 0.17                         & 6.96 $\pm$ 0.16                         \\
$\textrm{DFS}_{\textrm{SURE}}$        & 0.85 $\pm$ 0.03                         & 1.11 $\pm$ 0.08                         & 2.26 $\pm$ 0.16                         & 5.55 $\pm$ 0.14                        & 1.02 $\pm$ 0.07                         & 1.65 $\pm$ 0.18                         & 3.4 $\pm$ 0.15                          & 6.95 $\pm$ 0.16                         \\ \hline
\end{tabular}
\end{table*}

We illustrate denoising performance with the SURE Monte-Carlo estimator on two relatively small graphs. This allows for the explicit eigendecomposition of their associated Laplacian matrices and computation of the SGWT and SURE with which we compare our proposed method.

\subsubsection{New York City taxis}

Considering the number of taxi pickups at each intersection from the last experiment as our signal, we now apply the differential privacy mechanisms presented in Section \ref{sec:privacy} to sanitize it. Both the classical and analytic Gaussian mechanisms are used for different values of the privacy budget $\varepsilon$ and therefore noise levels $\sigma$, while the other privacy parameter is set to $\delta = 10^{-6}$ as in \cite{balle2018}. Note that to satisfy the constraint associated with the maximum value taken by the $\varepsilon$ parameter, the resulting noise levels are particularly high. We aim to protect the taxi passengers and assume they only take a taxi once within an hour. This gives an upper bound of their individual contribution to the signal and thus we have an $l_2$-sensitivity of $\Delta = 1$.

We compare different denoising methods using signal-to-noise ratio $\textrm{SNR}(f, \hat f) = 20 \log_{10}(\| f \|_2 / \| f - \hat f \|_2)$ as a performance measure. We also compute it between the original and noisy signals to get a baseline of the amount of input noise after sanitization: $\textrm{SNR}_{\textrm{in}} = \textrm{SNR}(f, \tilde f)$. Three denoising methods based on the application of a level-dependent James-Stein thresholding function to wavelet coefficients are considered, each of them uses a different criterion to select the optimal thresholds: (1) an oracle estimator that directly computes the SGWT and minimizes the real MSE; (2) a second estimator that instead minimizes SURE; and (3) our proposed estimator that approximates the SGWT with Chebyshev-Jackson polynomials and estimates SURE with Monte-Carlo. As shown by \cite{donoho1995}, for a coordinate-wise thresholding process such as James-Stein, SURE reaches its minimum for some threshold $t$ chosen among the absolute values of the noisy wavelet coefficients $\{ |\widetilde F_i|, i=1, \ldots, n(J+1)\}$. We further reduce this set to its percentiles to find a compromise between the range and number of candidate threshold values.

These estimators are compared to the DFS fused lasso, a regularization method introduced in \cite{padilla_2018}. It first performs a standard depth-first search (DFS) traversal algorithm to reduce the initial graph to a chain graph. Then, it runs a 1-dimensional fused lasso \cite{tibshirani2005}, a special case of graph trend filtering \cite{Wang2016}, over this simpler graph. In doing so, this method avoids the prohibitive computational cost of standard graph trend filtering over an arbitrary graph at the expense of less statistical accuracy. Here, the comparison is made on unweighted graphs as the DFS fused lasso is limited to them, whereas the SGWT can be applied to graphs with edge weights. In the experiments, the DFS and fused lasso are respectively conducted with the \pkg{igraph} \cite{igraph} and \pkg{glmgen} \cite{glmgen} R packages.

Table \ref{table:ny} summarizes the results of this experiment over 10 sanitization realizations. We observe that the wavelet transform oracle estimator ($\textrm{SGWT}_{\textrm{MSE}}$) performs better than the oracle DFS fused lasso ($\textrm{DFS}_{\textrm{MSE}}$) for most values of privacy budget. When considering stronger degrees of privacy with the classical Gaussian mechanism which requires the most amount of input noise, the oracle DFS fused lasso presents better results. Our approach combining Chebyshev-Jackson polynomial approximations with SURE Monte-Carlo estimation ($\textrm{SGWT}_{\textrm{SURE, MC}}^{\textrm{CJ}}$) gives slightly lower SNR values than its oracle counterpart but nevertheless shows better denoising performance compared to the regularization method except for the case where the noise is very high.

\subsubsection{San Francisco Taxis}

We check these initial results with a second dataset that contains the GPS coordinates of 536 taxis collected over a month in the San Francisco Bay Area \cite{Piorkowski2009CRAWDAD}. Each entry consists of the taxi location and whether it currently has passengers at a given time with approximately one minute between updates. In a similar fashion as for the previous experiment, we concentrate on the city of San Francisco and get the associated graph of the street network from \pkg{OSMnx}. It is about twice as large with 9573 nodes and 15716 edges, causing a longer but still practicable computation of the SGWT.

Pickup locations are inferred by keeping the entries whose occupancy status goes from "free" to "occupied", giving an approximation close to the minute. We build a signal by counting these pickups projected to the nearest intersection on the day of May 25, 2008. Sanitization is then applied with the analytic Gaussian mechanism and parameter values $\delta = 10^{-6}$ and $\Delta = 2$. The latter is chosen by assuming the individual passengers do a maximum of four taxi trips within a day, all starting from distinct places.

\begin{table}[h!]
\centering
\caption{Average SNR performance over 10 realizations of high to low privacy budget sanitization on the San Francisco graph.}
\label{table:sf}
\begin{tabular}{lccc}
\hline
$\varepsilon$           & 0.20                                   & 0.50                                    & 1                                       \\
$\sigma$             & 37.98                                  & 16.12                                   & 8.45                                    \\
$\textrm{SNR}_{\textrm{in}}$           & -11.95 $\pm$ 0.05                      & -4.51 $\pm$ 0.05                        & 1.1 $\pm$ 0.05                          \\ \hline
$\textrm{SGWT}_{\textrm{MSE}}$        & \gg 0.80 $\pm$ 0.27 &\gg 4.34 $\pm$ 0.11 &\gg 8.24 $\pm$ 0.07 \\
$\textrm{SGWT}_{\textrm{SURE}}$       & 0.77 $\pm$ 0.31                        & 4.32 $\pm$ 0.09                         & 8.22 $\pm$ 0.06                         \\
$\textrm{SGWT}_{\textrm{SURE, MC}}^{\textrm{CJ}}$ & 0.76 $\pm$ 0.32                        & 4.34 $\pm$ 0.12                         & 8.22 $\pm$ 0.08                         \\
$\textrm{DFS}_{\textrm{MSE}}$         & 0.22 $\pm$ 0.02                        & 3.02 $\pm$ 0.1                          & 6.77 $\pm$ 0.1                          \\
$\textrm{DFS}_{\textrm{SURE}}$        & 0.17 $\pm$ 0.07                        & 2.99 $\pm$ 0.12                         & 6.76 $\pm$ 0.1                          \\ \hline
\end{tabular}
\end{table}

Results presented in Table \ref{table:sf} are in line with those obtained above. The wavelet transform oracle estimator gives the best overall results and our method performs again better than the oracle DFS fused lasso for the considered privacy budget values.

\subsection{Denoising of Large Graph Signals}

In this experiment, we apply our method on a large graph whose scale prevents us from decomposing the Laplacian matrix due to the prohibitive computational cost. The road network of Pennsylvania from \cite{Leskovec2009} is such a graph consisting of 1088092 nodes and 1541898 edges.
Synthetic signals are generated on this graph following the methodology proposed in \cite{Behjat2016}: with two parameters $p \in (0,1)$ and $k \in \mathbb N$, we produce a signal $f_{p, k} = W^k x_p$ where $x_p$ is an i.i.d. realization of $n$ Bernoulli random variables of parameter $p$. As this data is entirely simulated and not related to the information of real individuals, the added noise does not depend on some privacy budget $\varepsilon$ value and is instead directly chosen. Here, a noisy signal $\tilde{f} = f_{0.001, 4} + \mathcal{N}(0, \sigma^2 I_n)$ is generated for different values of $\sigma$.

\begin{figure}[ht]
    \centering
    \includegraphics[width=0.48\textwidth]{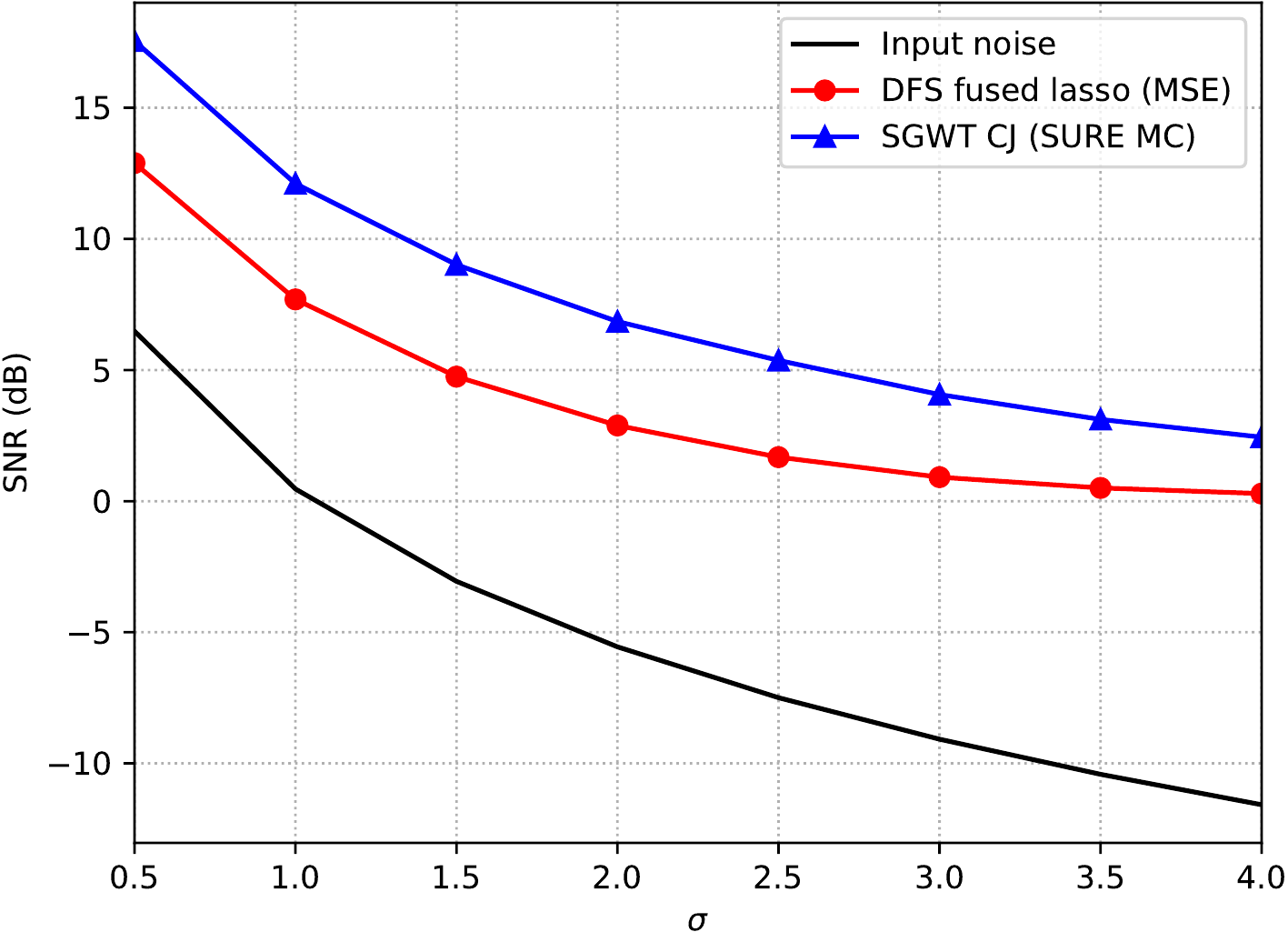}
    \caption{Average SNR performance over 5 realizations of each noise level setting on the Pennsylvania graph}
    \label{fig:snr_pa}
\end{figure}

Figure \ref{fig:snr_pa} presents the average SNR values over 5 noise realizations for our proposed estimator and the oracle DFS fused lasso. We see that the results observed on small graph signals extend well to the large-scale setting, with better performance for the Chebyshev-Jackson polynomial approximation with the SURE Monte-Carlo estimator on a range of input SNR similar to the previous experiments.

Regarding computing time, the DFS fused lasso is, however, more efficient than the approximated SGWT and estimated SURE for this application. On a standard laptop (Intel Core i5@1.70GHz-16Go DDR4@2400MHz), each of the realizations is denoised in 4 seconds by the regularization method while it takes less about 3 minutes for the latter after an initial estimation of the SURE weights done in 1 minute. We do not observe a significant difference between drawing Monte-Carlo samples from the Rademacher or the Gaussian distributions, both take the same amount of time. In our procedure, the most time-consuming step is the threshold optimization (2m30s), followed by the inverse wavelet transform (25s) and the forward transform (7s).

\section{Conclusion and Perspectives}

In this paper, we propose an extension of SURE to large-scale graphs in the context of signal denoising with thresholding of SGWT coefficients. In particular, we use Monte-Carlo and Chebyshev-Jackson polynomial approximation to build an estimator of its weights in order to avoid the computationally expensive eigendecomposition of the graph Laplacian matrix. Provided expressions for the variance of both weights and SURE estimators show that the Rademacher distribution is better suited than the Gaussian one for this method. We evaluate our data-driven approach through numerical experiments with an application in differential privacy to improve the utility of sanitized graph signals. Results show the MSE can be efficiently estimated with our extended SURE on small and large graphs. Additionally, this methodology shows better performance than the DFS fused lasso.

There is room for improvement in this approach to remove noise with better precision. For instance, the thresholding function we used can be generalized to $\tau(x, t) = x \max \{ 1 - t^\beta |x|^{-\beta}, 0 \}$ with $\beta \geq 1$. Common choices for $\beta$ include soft thresholding ($\beta = 1$) and hard thresholding ($\beta = \infty$) but an optimization algorithm for this parameter would be more beneficial to further improve performance. An additional thresholding strategy worth considering is block thresholding which partitions wavelet coefficients within each scale to identify localized features in the signal. Depending on the regularity of the original signal, this can help to remove noise with more accuracy as a different threshold value is selected for each block. An expression of SURE provided for block thresholding processes with SGWT coefficients by \cite{de2019data} could be extended to large graphs with our approach.

Another direction for future research is to adapt our methodology to be run on a distributed system in order to further reduce computing time. First, sanitization with the Gaussian mechanism only consists of the addition of independent Gaussian noise to each node of the graph. Differential privacy in this case can thus be achieved in a distributed manner over subgraphs of the initial graph. Threshold selection by SURE optimization can also be computed in a distributed manner thanks to the additive nature of the SURE formula. But the Laplacian matrix and the SGWT cannot be directly decomposed over separated groups of graph nodes. This yields at least two important challenges in order to distribute 1) the computation of the weights in the SURE formula, and 2) the SGWT thresholding procedure. For specific graph structures (e.g. relatively distinct subgraphs), localization properties of the SGWT would certainly help finding accurate approximations for these distributed computations.


\appendices

\section{Computation of $\mathbf V[\hatcov_{ij}^2]$}
\label{appendix:gamma_variance}
The formula of the weights $\hatcov^2_{ij}$ is given by
\begin{equation*}
    \hatcov^2_{ij} = \frac1N \sum_{k=1}^N \left ( \sum_{p=1}^n \WT_{ip} \varepsilon_{pk} \right ) \left ( \sum_{p=1}^n \WT_{jp} \varepsilon_{pk} \right).
\end{equation*}

The variance of $\hatcov^2_{ij}$ reads
\begin{equation*}
  \mathbf V[\hatcov^2_{ij}]=\frac1N \mathbf V \left [ \left ( \sum_{p=1}^n \WT_{ip} \varepsilon_{p1} \right ) \left ( \sum_{p=1}^n \WT_{jp} \varepsilon_{p1} \right ) \right ].
\end{equation*}

Then, on the one hand
\begin{align*}
  \mathbf E & \left [ \left ( \sum_{p=1}^n \WT_{ip} \varepsilon_{p1} \right )^2 \left ( \sum_{p=1}^n \WT_{jp} \varepsilon_{p1} \right )^2 \right ] \\
  & = \sum_{p,q,r,s=1}^n \WT_{ip} \WT_{iq} \WT_{jr} \WT_{js} \mathbf E[\varepsilon_{p1}\varepsilon_{q1} \varepsilon_{r1} \varepsilon_{s1}] \\
  & = \mathbf E[\varepsilon_{11}^4] \sum_{p=1}^n \WT_{ip}^2 \WT_{jp}^2 + \mathbf E[\varepsilon_{11}^2]^2 \sum_{\substack{p,r=1, \\ p \neq r}}^n \WT_{ip}^2\WT_{jr}^2 \\
  & + 2\mathbf E[\varepsilon_{11}^2]^2 \sum_{\substack{p,q=1, \\ p \neq q}}^n \WT_{ip} \WT_{iq} \WT_{jp} \WT_{jq}.
\end{align*}

On the other hand,
\begin{align*}
  \mathbf E \left [ \left ( \sum_{p=1}^n \WT_{ip} \varepsilon_{p1} \right )^2 \right ] & = \sum_{p,q=1}^n \WT_{ip} \WT_{iq} \mathbf E[\varepsilon_{p1} \varepsilon_{q1}]\\
  & = \mathbf E[\varepsilon_{11}^2] \sum_{p=1}^n \WT_{ip}^2.
\end{align*}

Finally,
\begin{align*}
  \mathbf V & \left [ \left ( \sum_{p=1}^n \WT_{ip} \varepsilon_{p1} \right ) \left ( \sum_{p=1}^n \WT_{jp} \varepsilon_{p1} \right ) \right ] \\
  & = \mathbf E[\varepsilon_{11}^4] \sum_{p=1}^n \WT_{ip}^2 \WT_{jp}^2 + \mathbf E[\varepsilon_{11}^2]^2 \sum_{\substack{p,r=1, \\ p \neq r}}^n \WT_{ip}^2\WT_{jr}^2 \\
  & + 2 \mathbf E[\varepsilon_{11}^2]^2 \sum_{\substack{p,q=1, \\ p \neq q}}^n \WT_{ip} \WT_{iq} \WT_{jp} \WT_{jq} \\ 
  & - \mathbf E[\varepsilon_{11}^2]^2 \sum_{p,q=1}^n \WT_{ip}^2 \WT_{jq}^2 \\
  & = \mathbf V[\varepsilon_{11}^2] \sum_{p=1}^n \WT_{ip}^2 \WT_{jp}^2 \\
  & + 2\mathbf E[\varepsilon_{11}^2]^2 \sum_{\substack{p,q=1,\\ p \neq q}}^n \WT_{ip} \WT_{iq} \WT_{jp} \WT_{jq}.
\end{align*}

\section{Computation of $\mathbf V [\mathbf{\widehat{SURE}}(h) | \widetilde F]$}
\label{appendix:sure_variance}

From Equation~\eqref{eq:sureh}, it follows that
\begin{align*}
  \mathbf V & [\mathbf{\widehat{SURE}}(h) | \widetilde F]  = 4 \sigma^4 \mathbf E \left [ \left ( \sum_{i,j=1}^{n(J+1)} (\hatcov_{ij}^2 - \gamma_{ij}^2) \partial_jh_i(\widetilde F) \right )^2 \right ] \\
  &= 4 \sigma^4 \sum_{i,j,k,\ell=1}^{n(J+1)} \partial_j h_i(\widetilde F) \partial_k h_\ell(\widetilde F) \mathbf E[(\hatcov_{ij}^2-\gamma_{ij}^2)(\hatcov_{k\ell}^2-\gamma_{k\ell}^2)] \\
   &=4 \sigma^4 \sum_{i,j,k,\ell=1}^{n(J+1)} \partial_j h_i(\widetilde F) \partial_k h_\ell(\widetilde F) \left [ \mathbf E[\hatcov_{ij}^2 \hatcov_{k\ell}^2] - \gamma_{ij}^2 \gamma_{k\ell}^2 \right ].
\end{align*}

Then,
\begin{align*}
  N^2\mathbf E[\hatcov_{ij}^2 \hatcov_{k\ell}^2] = \sum_{a,b=1}^N \mathbf E \Bigg [ & \left ( \sum_{p=1}^n \WT_{ip} \varepsilon_{pa} \right ) \left ( \sum_{p=1}^n \WT_{jp} \varepsilon_{pa} \right ) \\
     & \left ( \sum_{p=1}^n \WT_{kp} \varepsilon_{pb} \right )\left ( \sum_{p=1}^n \WT_{\ell p} \varepsilon_{pb} \right ) \Bigg ].
\end{align*}

Developing each term of the sum above, it follows
\begin{multline*}
  N^2 \mathbf E[\hatcov_{ij}^2 \hatcov_{k\ell}^2] \\
  = \sum_{a,b=1}^N \sum_{p,q,r,s=1}^{n} \WT_{ip} \WT_{jq} \WT_{kr} \WT_{\ell s} \mathbf E[\varepsilon_{pa}\varepsilon_{qa}\varepsilon_{rb} \varepsilon_{sb}].
\end{multline*}

Thus, on the one hand
\begin{align*}
  N^2 \mathbf E[\hatcov_{ij}^2 \hatcov_{k\ell}^2] & = N \mathbf E[\varepsilon_{11}^4]\sum_{p=1}^n \WT_{ip} \WT_{jp} \WT_{kp} \WT_{\ell p} \\
  & + N \mathbf E[\varepsilon_{11}^2]^2 \sum_{\substack{p,q=1, \\ p \neq q}}^{n} \WT_{ip} \WT_{jp} \WT_{kq}\WT_{\ell q} \\
  & + N \mathbf E[\varepsilon_{11}^2]^2 \sum_{\substack{p,q=1, \\ p \neq q}}^{n} \WT_{ip} \WT_{jq} \WT_{kp}\WT_{\ell q} \\
  & + N \mathbf E[\varepsilon_{11}^2]^2 \sum_{\substack{p,q=1, \\ p \neq q}}^{n} \WT_{ip} \WT_{jq} \WT_{kq}\WT_{\ell p} \\
  & + N(N-1) \mathbf E[\varepsilon_{11}^2]^2 \sum_{p,q=1}^{n} \WT_{ip} \WT_{jp} \WT_{kq}\WT_{\ell q}.
\end{align*}

On the other hand,
\begin{align*}
  \gamma^2_{ij} \gamma^2_{k\ell} & = \mathbf E[\varepsilon_{11}^2]^2 \left ( \sum_{p=1}^n \WT_{ip} \WT_{jp} \right ) \left ( \sum_{p=1}^n \WT_{kp} \WT_{\ell p} \right ) \\
  & = \mathbf E[\varepsilon_{11}^2]^2 \sum_{p,q=1}^n \WT_{ip} \WT_{jp} \WT_{kq} \WT_{\ell q}. 
\end{align*}

Finally, the difference is given by
\begin{align*}
  N^2 \mathbf E & [\hatcov^2_{ij}\hatcov_{k\ell}^2]-N^2\gamma_{ij}^2\gamma_{k\ell}^2 \\
  & = N \mathbf E[\varepsilon_{11}^4]\sum_{p=1}^n \WT_{ip} \WT_{jp} \WT_{kp} \WT_{\ell p} \\
  & + N \mathbf E[\varepsilon_{11}^2]^2 \sum_{\substack{p,q=1, \\ p \neq q}}^{n} \WT_{ip} \WT_{jp} \WT_{kq}\WT_{\ell q} \\
  & + N \mathbf E[\varepsilon_{11}^2]^2 \sum_{\substack{p,q=1, \\ p \neq q}}^{n} \WT_{ip} \WT_{jq} \WT_{kp}\WT_{\ell q} \\
  & + N \mathbf E[\varepsilon_{11}^2]^2 \sum_{\substack{p,q=1, \\ p \neq q}}^{n} \WT_{ip} \WT_{jq} \WT_{kq}\WT_{\ell p} \\
  & + N(N-1) \mathbf E[\varepsilon_{11}^2]^2 \sum_{p,q=1}^{n} \WT_{ip} \WT_{jp} \WT_{kq}\WT_{\ell q} \\
  & - N^2 \mathbf E[\varepsilon_{11}^2]^2 \sum_{p,q=1}^n \WT_{ip} \WT_{jp} \WT_{kq} \WT_{\ell q}.
\end{align*}

Hence,
\begin{align*}
  N^2 \mathbf E & [\hatcov^2_{ij}\hatcov_{k\ell}^2]-N^2\gamma_{ij}^2\gamma_{k\ell}^2 \\
  & = N \mathbf V[\varepsilon_{11}^2] \sum_{p=1}^n \WT_{ip} \WT_{jp} \WT_{kp} \WT_{\ell p} \\
  & + N \mathbf E[\varepsilon_{11}^2]^2 \sum_{\substack{p,q=1, \\ p \neq q}}^{n} \WT_{ip} \WT_{jq} \WT_{kp}\WT_{\ell q} \\
  & + N \mathbf E[\varepsilon_{11}^2]^2 \sum_{\substack{p,q=1, \\ p \neq q}}^{n} \WT_{ip} \WT_{jq} \WT_{kq}\WT_{\ell p}.
\end{align*}

Summarizing,
\begin{align*}
  \mathbf V[\mathbf{\widehat{SURE}}(h) | \widetilde F] & = \frac{4 \sigma^4}{N} \sum_{i,j,k,\ell=1}^{n(J+1)} \partial_j h_i(\widetilde F) \partial_k h_\ell(\widetilde F) \Bigg \{ \\
  & \phantom{+~} \mathbf V[\varepsilon_{11}^2] \sum_{p=1}^n \WT_{ip} \WT_{jp} \WT_{kp} \WT_{\ell p} \\
  & + \mathbf E[\varepsilon_{11}^2]^2 \sum_{\substack{p,q=1, \\ p \neq q}}^{n} \WT_{ip} \WT_{jq} \WT_{kp}\WT_{\ell q} \\
  & + \mathbf E[\varepsilon_{11}^2]^2 \sum_{\substack{p,q=1, \\ p \neq q}}^{n} \WT_{ip} \WT_{jq} \WT_{kq}\WT_{\ell p} \Bigg \}.
\end{align*}

\section*{Acknowledgment}
The authors are very grateful to the Associate Editor and the two anonymous referees for their thorough and useful comments.

\bibliographystyle{abbrv}
\bibliography{biblio}

\begin{IEEEbiographynophoto}{Elie Chedemail}
Elie Chedemail received the B.Sc. degree in mathematics and economics from the University of Rennes 1, Rennes, France in 2016 and the M.Sc. degree in statistics from the National School for Statistics and Data Analysis (ENSAI), Bruz, France in 2019. He is currently working towards the Ph.D. degree at ENSAI and Orange Labs, Cesson-Sévigné, France. His research interests include machine learning, graph signal processing and differential privacy.
\end{IEEEbiographynophoto}

\begin{IEEEbiographynophoto}{Basile de Loynes}
Basile de Loynes received a Ph.D. degree in fundamental mathematics from the University of Rennes (France) in 2012. From 2012 to 2013, he was a postdoctoral researcher with the Department of Mathematics at the University of Neuchâtel (Switzerland). From 2013 to 2014, he was an assistant professor at the University of Burgundy (France) and from 2014 to 2016 he was an assistant professor at the University of Strasbourg (France). He is currently an associate professor at CREST-Ensai. His research interests include random walks on graphs, long memory stochastic processes, functional analysis and graph signal processing.
\end{IEEEbiographynophoto}

\begin{IEEEbiographynophoto}{Fabien Navarro}
received the B.Sc., M.Sc. and Ph.D. degrees in Applied Mathematics from the University of Caen, Caen, France, in 2008, 2010 and 2013, respectively.  From 2014 to 2015, he was a Research Assistant Professor with the department of Mathematics and Statistics, Concordia University, Montreal, Canada. From 2015 to 2021, he was an Assistant Professor with the Center for Research in Economics and Statistic, Ecole Nationale de la Statistique et de l'Analyse de l'Information, Bruz, France.  He is currently an Associate Professor with the University of Paris 1 Panth\'eon-Sorbonne, Paris, France.  His research interests include nonparametric statistics, inverse problems, computational harmonic analysis, sparse representations, machine learning and statistical approaches in graph signal processing.
\end{IEEEbiographynophoto}

\begin{IEEEbiographynophoto}{Baptiste Olivier}
Baptiste Olivier received a Ph.D. degree in fundamental mathematics from the University of Rennes (France) in 2013. From 2013 to 2015, he was a postdoctoral researcher with the Department of Mathematics at the TECHNION in Haifa (Israel). From 2016 to 2021, he was a research and data scientist at Orange in Rennes (France). He is currently a senior data scientist at Ericsson in Stockholm (Sweden). His research interests include property (T), functional analysis, machine learning, graph signal processing and differential privacy.
\end{IEEEbiographynophoto}

\end{document}